\documentstyle[prl,aps,multicol,epsfig]{revtex}
\begin{document}
\draft
\widetext
\title{Point-contact spectroscopy  of MgB$_2$}
\author{  P.  Samuely,$^{1}$  P.  Szab\'o,$^{1}$ J.  Ka\v cmar\v
c\'{\i}k,$^{1}$ T.  Klein,
$^{2}$ A. G. M. Jansen$^{3}$  }
\address{$^1$Institute  of  Experimental  Physics,  Slovak
Academy of Sciences, Watsonova 47, SK-043 53 Ko\v{s}ice, Slovakia.}
\address{$^2$ Universit\' e  Joseph  Fourier,  BP  53,  F-38041
Grenoble Cedex 9, France.}
\address{$^3$ Grenoble High Magnetic Field Laboratory,
MPI-FKF and CNRS, F-38042 Grenoble, France.}
\date{\today}
\maketitle
\widetext
\begin{abstract}
\vspace{.5cm}

Point-contact spectroscopy measurements  on magnesium diboride
reveal  the  existence  of  two  superconducting energy gaps
closing at the same transition  temperature in line with the
multiband model  of superconductivity. The sizes  of the two
gaps  ($\Delta_{\pi}$  =  2.8  meV  and  $\Delta_{\sigma}$ =
6.5-7  meV) are  respectively  smaller  and larger  than the
expected weak coupling value  of the one-gap superconductor.
The  smaller  gap   is  rapidly filled by
a small magnetic field of about  1-2 Tesla much lower than the
real upper critical field $H_{c2}$. The larger gap is closed
at  the   anisotropic  $H_{c2}$.  Above   the  gap  energies
reproducible    non-linearities   are    observed   at    the
characteristic phonon energies of MgB$_2$.
\vspace{1cm}
\end{abstract}
\begin{multicols}{2}
\narrowtext
\section{Introduction}
Point-contact spectroscopy  offers the possibility  to study
very  fundamental  superconducting  properties.  Namely, the
superconducting energy gap can  be addressed via the Andreev
reflection and  the electron-phonon (boson)  interaction can
be  inferred  as  well.  From  the  very beginning since the
discovery  of  MgB$_2$  \cite{nagamatsu}  this technique has
been applied  for the investigation of  this compound. Here,
we  shortly  review  the  state  of  art which unequivocally
supports the  fact that MgB$_2$  represents an extraordinary
example  of  multigap   superconductivity  which  phenomenon
attains interest since the fifties \cite{suhl,binning}.

The   electronic  structure   calculations  have   predicted
\cite{liu}  that several  electronic bands  cross the  Fermi
level in MgB$_2$ with  two different superconducting gaps at
different sheets. These predictions have been experimentally
evidenced  by   different  techniques  like   specific  heat
measurements             \cite{bouquet},            tunneling
\cite{giubileo,iavarone,suderow},  Raman  spectroscopy  \cite{chen},
and   so  on.   Among  these   investigation  methods   also
spectroscopy  based on  the Andreev  reflection process gave
one  of  the  proofs  of  such  multi-gap  superconductivity
\cite{szabo}. For
the  larger   gap  $\Delta  _{\sigma}$   attributed  to  the
two-dimensional  $\sigma$-band  parallel   to  the  $c$-axis
originating from  the boron $p_{x-y}$  orbitals, the reduced
gap  value $2\Delta  _{\sigma}  /k_BT_c  \simeq 4$  has been
found. The smaller gap $\Delta  _{\pi}$ on the 3D $\pi$-band
of the boron $p_z$-orbitals has the reduced value much below
the  BCS weak  coupling limit  of a  one-band superconductor
($2\Delta _{\pi} /k_BT_c \simeq 1.7  $). The behavior of the
both  gaps  in  magnetic  field  has  been studied in detail
\cite{szabo,bouquet2,eskildsen}. The
small  gap is filled  very rapidly by a small field
of  about   1-2  Tesla  at  low   temperatures.
The   large  gap  is  closed  at  the real upper
critical field of MgB$_2$ which is anisotropic.

The  relatively high  value of  the superconducting critical
temperature $T_c$  in MgB$_2$ is supposed  to originate from
the strong  anharmonic coupling of the  $\sigma$-band to the
E$_{2g}$ phonons  \cite{liu}. The recent  de Haas-van Alphen
experiments  \cite{yelland}  give  indications  for  such an
anisotropic  electron-phonon  interaction,  but  there is no
direct spectroscopic  evidence for the overall  shape of the
Eliashberg function of  the electron-phonon interaction. The
very  first   point-contact  spectroscopy  studies   of  the
electron-phonon    interaction     which    are    available
\cite{bobrov,D'yachenko,Szabo02,Yanson3} indicate
reproducible   non-linearities    of   the   current-voltage
characteristics   at  the   MgB$_2$  characteristic   phonon
energies.  Further experiments  and theoretical  studies are
necessary to  elucidate details of  the phononic interaction
mechanism.

\section{Superconducting energy gaps}

Below $T_c$ a phase coherent state of Cooper pairs is formed
in  a superconductor.  Transport of  charge carriers  across
a normal-metal/superconductor  (N/S) interface  involves the
unique process  of Andreev reflection. If  the N/S interface
consists  of a  ballistic point  contact with  the mean free
path $l$  bigger than the  diameter of the  contact orifice,
the excitation energy $eV$  of charge carriers is controlled
by the applied voltage $V$. For a quasiparticle incident on
the N/S  interface with an excitation  energy $eV < \Delta$,
a direct  transfer  of  the  charge  carriers  is  forbidden
because of  the existence of the  energy gap $\Delta$ in the
quasiparticle  spectrum  of  the  superconductor.  For this
case,  the charge  transfer  of  an injected  charge carrier
takes place via the retroreflection  of a hole back into the
normal  metal with  the formation  of a  Cooper pair  in the
superconductor.  At   excitation  energies  above   the  gap
quasiparticles  can  be   transferred  directly  across  the
interface.  The Andreev  reflection process  leads to  a two
times higher conductance of a  N/S contact at $V < \Delta/e$
(zero-temperature limit) for  the  case  of  ballistic  transport  with  transmission
probability  of the  charge carriers  $T =  1$. The opposite
limit  of  ballistic  transport  is  given  by the tunneling
process for an insulating barrier  at the interface where $T
<< 1$.  For Giaever tunneling  between a superconductor  and
a normal metal,  it is well known  that the conductance goes
to  zero for  $eV <  \Delta$ (again  in the zero-temperature
limit). In contrast to the point-contact
tunneling where the well-developed gap structures could also
be  due  to  other  effects,  like  small  particle charging
\cite{mullen},  the Andreev  reflection directly  probes the
coherent  superconducting state.  The more  general case for
arbitrary  transmission  $T$  has  been  treated by Blonder,
Tinkham  and  Klapwijk  (BTK)  \cite{Blonder82}. The voltage
dependence of the conductance of  a N/S contact gives direct
spectroscopic  information  on   the  superconducting  order
parameter  $\Delta$. The  conductance data  can be  compared
with the BTK theory using as input parameters the energy gap
$\Delta$, the parameter $Z$ (measure for the strength of the
interface  barrier   with  transmission  coefficient   $T  =
1/(1+Z^2)$ in  the normal state),  and a parameter  $\Gamma$
for the quasi-particle  lifetime broadening \cite{plecenik}.
The evolution  of the $dI/dV$  vs. $V$ curves  for different
interfaces  characterized with  the barrier  strength $Z$ is
schematically presented in Fig. 1a.

The  first  point-contact  measurements  on sintered MgB$_2$
pellets  were performed  by Schmidt  et al.  \cite{schmidt}.
They showed  that by the  Au tip it  was possible to  change
from  the  Giaever  tunneling  regime  to  the  pure Andreev
reflection. In both cases the $s$-wave gap was observed with
a value of 4.3 meV. The  smallness of this value compared to
the expected value for a 39-K BCS superconductor was ascribed
to a chemically modified surface layer with lower $T_c$, but
the  particular   $T_c$  of  the   point  contact  was   not
determined.

Kohen and Deutscher \cite{kohen} found the energy gap in the
polycrystalline MgB$_2$  scattered between 3 and  4 meV. The
point contacts were done by means of the Au tip. The
temperature dependence of the 3  meV energy gap followed the
BCS form with $T_c$ about 29 K. But even for this transition
temperature, reduced  in comparison with  the bulk material,
the resulting $2\Delta/k_BT_c \simeq 2.4 $ is much too small
for a  weak coupling BCS superconductor.  The Fermi velocity
$4.7 \times 10^7$~cm/s was calculated from the $Z$ parameter
with an assumption that the lowest  $Z$ originated just from the Fermi velocity mismatch between
Au and MgB$_2$. The obtained value was
in   agreement   with   the   band   structure  calculations
\cite{kortus}.

Measurements   by  Plecenik   et  al.   \cite{plecenik1}  on
MgB$_2$/Ag  and  MgB$_2$/In  junctions  performed on MgB$_2$
wires indicated  the same problem  with a small  size of the
energy gap.  The temperature dependence of  the 4~meV energy
gap showed  a significant deviation from  the BCS form which
has been  ascribed to an  interference of two  gaps with two
different  $T_c$'s  in  line  with  the  published  multigap
scenario \cite{liu}. In their model Liu et al. \cite{liu}
suggested  that  the  two  gaps  are  sensitive  to impurity
scattering. In contrast to  Anderson's theorem for a one-gap
superconductor,  non-magnetic  interband
scattering  can decrease  $T_c$  in  this case  of multi-gap
superconductivity.
Moreover,  in the  limit of
strong scattering  both gaps will be averaged  into one single
BCS gap ($\Delta  (0) \simeq  4$~meV)  which will  close at
$T_c  \simeq 27$~K. Plecenik  et al.  suggested that  their
point  contacts  represent  a  parallel  connection  of  two
junctions: one probing the small gap in the clean region and
another  the  averaged  BCS  gap  in  the  dirty  part  with
a reduced $T_c$.

Laube   et   al.   \cite{laube}    and   Gonnelli   et   al.
\cite{gonnelli}  have  also  proposed explanations for the
observed small gaps in their data in terms of
multigap superconductivity.

Indications  for the two-gap superconductivity have  been found in the
point-contact measurements  of S. Lee et  al. \cite{lee} and Li
et al.  \cite{li}. The latter  group claim the  second large
gap being around 10 meV which  is much higher than the other
experimental observations and  the theoretical predictions.

In our work  \cite{szabo} we have shown the  presence of two
superconducting energy  gaps in MgB$_2$ closing  at the same
bulk  $T_c$.   The  point-contact  measurements   have  been
performed on  polycrystalline MgB$_2$ samples  with critical
temperature $T_c = 39.3$ K and width $\Delta T_c = 0.6$ K of
the  superconducting transition.  The pressure-type contacts
could be  adjusted in-situ by  pressing a copper  tip on the
freshly polished surface of the MgB$_2$ superconductor using
a differential  screw  mechanism.  The  crystallites  of the
MgB$_2$ samples were larger than a few micron. For a typical
size  of  the  metallic   constriction  below  10~nm,  the
point-contact   probes   therefore   presumably   a   single
crystallite but of unknown orientation.

In  figure 2  we show  typical examples  of the differential
conductance  $dI/dV$ versus  voltage spectra  for Cu-MgB$_2$
point contacts in the superconducting  state at 4.2~K. The three
upper curves reveal a  clear two-gap structure corresponding
to  the  maxima  placed   symmetrically  around  zero  bias.
However, 90\%  of the contacts reveal  a spectrum dominantly
showing   the  small-gap   structure  with   only  a   small
shoulder-like  contribution  around  the  large-gap voltage.
These  variations from  contact to  contact are  ascribed to
current  injection  along  different  crystal  directions of
locally probed crystallites.
The conductance curves  of figure 2 have been  fitted to the
BTK  theory  using  the  sum  of  two contributions, $\alpha
\Sigma_{\pi  } +  (1-\alpha)\Sigma_{\sigma}$, with  a weight
factor   $\alpha$   between   the   small-   and   large-gap
contributions,    $\Sigma_{\pi}$    and    $\Sigma_{\sigma}$
respectively.  Details  of the  fitting  to  the  sum of two BTK
conductances  are  shown  in  Fig.  1b. The weight factor $\alpha$
varied from 0.65 to 0.95 for the different contacts.
The resulting energy gaps are $\Delta_{\pi}=2.8 \pm 0.1$~meV for
the small  gap and $\Delta_{\sigma}=6.8 \pm  0.3$~meV for the large
gap.

Point-contact data  with a very low peak(s)  height to background
ratio  (PHBR)  have  been  reported  in  some  of  the above
mentioned papers.  In the case  of the (prevailing)  one-gap
spectrum the low PHBR (e.g. in comparison with those in Fig.
1a) is caused by the  large smearing parameter $\Gamma$. But
even a structureless, parallel  leakage current to the normal
regions on the  surface of MgB$2$ can occur  as mentioned in
\cite{plecenik1}. In  the case of  an important contribution
of  the both  conductances to  the spectrum  (we have  found
$\alpha = 0.65$)  with   generally  two  different  $\Gamma$
broadenings and $Z$ parameters, PHBR is more complicated but
for  a low  $\Gamma \rightarrow  0$ it never  went below  1.4 in  our
measurements.  The  spectra  shown  in  Fig.  2 reveals also
the lowered PHBR  in comparison with  those the Fig.  1b and Fig.
3.  Again, it  is  caused  by the  increased  broadening parameter. For
example           $\Gamma_{\pi}=0.13\Delta_{\pi}$          and
$\Gamma_{\sigma}=0.07\Delta_{\sigma}$ is found for  the top
curve  and   $\Gamma_{\pi}=0.22\Delta_{\pi}$
 in the case of the bottom curve.

In  the   calculations  of  two-band   superconductivity  in
MgB$_2$ \cite{liu}, the small gap is  situated on the
isotropic   $\pi$   band   and   the   large   gap   on  the
two-dimensional $\sigma$ band (cylindrically shaped parallel
to the $c$-axis).  The Fermi-surface topology  of the two  bands
would explain  the observed differences  in the spectra  for
different  contact   orientations.  The  small   gap,  being
isotropic, will  be observed for  all directions of  current
injection.  However, the  two-dimensional cylindrical  Fermi
surface will give most contribution to the interface current
for  current   injection  parallel  to   the  ab-plane.  The
observation of  two-gap structures in  a minority of the
investigated  contacts might   mean that  the crystallites in
the  polycrystalline samples  are mostly  textured with  the
$ab$-planes  at the  surface. Thus,  the spectra  in Fig.  2
represent an  evolution from the  $c$-axis tunneling to  the
$\pi$-band   (bottom   curve)   to   the  bigger  $ab$-plane
contributions (higher curves) revealing the both $\pi$- and
$\sigma$-gaps.

The  smaller gap  could   $a priori$  be caused by  different
reasons,    including   a    surface   layer    of   reduced
superconductivity or
surface  proximity
effects. However, most of  the scenarios would still require
a scaling  of   the  gap  with   the  critical  temperature.
That is  why it was  important to show  an existence of  the
small gap
at high temperatures in order to establish its origin in the
two-band superconductivity. The
temperature dependencies  of the point-contact  spectra have
been examined on different samples. Due to thermal smearing,
the two  well resolved peaks in  the spectrum merge together
as the temperature increases. Consequently, the presence of
two gaps  is not so evident  in the raw data  (see Fig. 3a).
For instance at 25 K the  spectrum is reduced to one smeared
maximum around zero bias. Such  a spectrum itself could
be  fitted by  the BTK  formula with  only one  gap, but the
transparency   coefficient   $Z$   would   have   to  change
significantly in comparison with  the lower temperatures and
moreover a  large smearing factor  $\Gamma$ would have to be
introduced.  However our  data could  be well  fitted at all
temperatures by  the sum of  two BTK contributions  with the
transparency coefficient $Z$ and  the weight factor $\alpha$
kept constant and without  any parallel leakage conductance.
The point-contact conductances of  one spectrum at different
temperatures  are  shown  in   Fig.  3a  together  with  the
corresponding   BTK   fits.   The   resulting   energy  gaps
$\Delta_{\sigma}$  and  $\Delta_{\pi}$  together  with those
obtained for two other  point contacts with different weight
values $\alpha$ are shown in Fig. 3b. In a classical
BCS theory,  an energy gap  with $\Delta =  2.8$~meV could
not exist for a system  with $T_c$ above $2\Delta/3.5k_{B} \sim$
19 K, but as shown in  Ref.\cite{ttpam} even a one-gap spectrum
(shown  as the  bottom curve  in  Fig.  2) leads  to the  same
temperature dependence $\Delta_{\pi}(T)$ with the bulk $T_c$ as those indicated
in Fig. 3b.

Since it is evident  that both gaps are closing
near  the same  bulk transition  temperature, our  data give
experimental  support for  the  two-gap model.  We obtained
a very  weakly coupled  gap with $2\Delta_{\pi}/k_BT_c\simeq
$ 1.7     and     a     strongly     coupled     gap    with
$2\Delta_{\sigma}/k_BT_c  \simeq$ 4  in very  good agreement
with   the   predictions   of    Brinkman   {\it   et   al.}
\cite{brinkman}   (a  3D   gap  ratio  $2\Delta_{\pi}/k_BT_c
\simeq $  1.61 and a  2D gap ratio  $2\Delta_{\sigma}/k_BT_c
\simeq$ 4.22).  The temperature dependencies  are in a  good
agreement  with  the  prediction  of  the  BCS theory. Small
deviations from this theory have  been predicted by Liu {\it
et al.} but  these deviations are within our  error bars for
the  large gap  $\Delta_{\sigma}$ while  in the  case of the
small gap $\Delta_{\pi}$ there is  a tendency for more rapid
closing at  higher temperatures near to  $T_c$ (see Fig. 3b)
as expected theoretically.

Later point-contact  measurements confirmed
the two-gap scenario in MgB$_2$.
Bugoslavsky et al. \cite{bugoslav} have demonstrated in their
point contacts made on  MgB$_2$ $c$-oriented thin films that
the two  distinct gaps can be evidenced already in  the raw data
at low temperatures  ($\Delta_1 = 2.3 $ meV  and $\Delta_2 =
6.5$ meV) and they both close at the same bulk $T_c$.
The observed  differences in relative weight of  the two gaps
have  also   been  ascribed  to   differences in crystallographic
orientation under the Au tip and in barrier strength
of their junctions.
Bobrov   et   al.   \cite{bobrov}   and   Naidyuk   et   al.
\cite{naidyuk} have presented the Andreev reflection spectra
on the  Ag/MgB$_2$ $c$-axis oriented film  point contacts at
4.2  K.  Their  histogram  of  the   gap distribution
reveals  distinct maxima  at 2.4  and 7  meV.

Very recently Gonnelli et al. \cite{gonnelli2} have presented the
point-contact  spectroscopy on the  MgB$_2$ single crystals.
Their   directional   measurements   have   shown   that  the
point-contact  current  parallel  to  the MgB$_2$ $ab$-plane
probes both the $\sigma$  and $\pi$ bands revealing the
two gaps  ($\Delta_{\sigma} = 7.1$  meV and $\Delta_{\pi}  =
2.9$ meV) while the point-contact current injected along the
$c$-axis reveals  only one gap  on the $\pi$-band.  They also
confirmed that  the temperature dependence  of the small  gap
reveals a  stronger suppression near  $T_c$ while the  large
gap closes in line with the BCS temperature dependence.

Figure 4  demonstrates that the application of  a magnetic field
has been important to  show the co-presence  of both small and large energy gap
at high temperatures   much    above   the   temperature    $T^*   =
2\Delta_{\pi}/3.52k_B \simeq 19 $~K corresponding the BCS weak
coupling  limit.  Indeed,  the  two  distinct  gaps  clearly
visible   in  the   Andreev  reflection   spectrum  at   low
temperatures  get completely  merged together  due to thermal
smearing at  temperatures above 20 K.  However, because the small
gap  is very  sensitive to  the applied  field, the large-gap spectrum
becomes  better resolved in a field of a few tenths
of Tesla even at temperatures above  30~K.
The suppression of the small gap structure enhances the
large-gap structure which is only suppressed at larger fields.
This  difference in the field sensitivity of  both  gaps
published  for the first  time  in  our  previous paper \cite{szabo}
opened the question about different  magnetic-field scales in
the separate $\pi$ and $\sigma$ bands.

\section{Upper critical magnetic fields}

Fig.  5   shows  the  behavior of the
junction with the prevailing one-gap spectrum (weight factor
of 0.95 of the $\pi$-band contribution)
   as a function
of  magnetic  field  at  4.2, 15 and 25~K.  The  effect of the applied
magnetic field is very unexpected. The small magnetic fields
below 2 T first shift  the peak position to higher voltages.
In the usual case of a
one-gap superconductor the application of magnetic field can
only  lead  to  a  shrinkage  of  the  distance  between  the
gap-like  peaks  in   the  point-contact/tunneling  spectrum
\cite{BPBO}.  This is a simple  consequence of  the
fact that  the magnetic pair-breaking  for increasing applied
field fills up forbidden states inside
the gap.   Obviously, the
effect demonstrated in Fig. 5 is due
to an interplay of the  two gaps which are inevitably always
present  together  in  any  spectrum  with  arbitrary weight
factor  $\alpha$.  As
calculated  by  Brinkman  et  al.  \cite{brinkman}  even for
strictly $c$-axis  tunneling there is  one percent tunneling
probability into the $\sigma$-band.  However, unlike the case
of tunneling spectroscopy with  a very dominant contribution
for perpendicular-to-the-barrier transport,  we have to note
that  the ballistic  transport in  metallic contacts injects
electrons   through  the   contact  with   a  much   larger
cos-$\theta$ angular spreading. Thus,  if in some cases only
a peak due to the   smaller  gap   is  apparent,   its  width   is  hiding
a contribution of the second gap.
Then, in a  magnetic field, the small gap  in the $\pi$-band
is  rapidly suppressed  and  the  large gap  will definitely
emerge. This  superposition of the  rapidly suppressing peak
of  the small  gap and  the emerging  of the  large one with
a very small weight factor causes  the observed shift of the
peak towards  higher voltages.  At even higher fields the
magnetic  pair-breaking   in  the  $\sigma$-band   leads  to
conventional  shrinkage of  the broadened  gap peaks  of the
large gap. Such  a smooth shift  without
any  major change  in peak  position during  transition from
prevailing observation of the small gap to the large one is a consequence of
an  important  broadening  already  at  zero magnetic field.
With  magnetic  field  the broadening  substantially increases and
the two "peaks" merge to the  one. Fig. 4 demonstrates the effect
of  the  field  on  the  spectrum  with a small
broadening  at  zero  field.  At  4.2  K  the  two  peaks or
shoulders from  the large and small  gaps are well separated
and  they just  shrink their  voltage position with a field.  At 20 K the
temperature smearing causes that the  two peaks merge to one
already at zero field.
Then, again the   applied   field   suppresses   the  small  gap
contribution,  the  peak  position  first  shifts  to higher
voltages and  at a certain field it starts to shrink.

Being  aware  of  possible  uncertainty  we  can  define the
magnetic field  where the shift  of the peak  towards higher
voltages turns back to lower voltages as the "upper critical
field"   $H_{c2}^{\pi}$  of   the  $\pi$-band   without  any
interband scattering.  At 4.2 K  this field is  at about 1.7
Tesla, at 15~K  at about 1.4~T, at 25 K  at about 0.7~T. The
full temperature  dependence of this  pair-breaking field of
the $\pi$-band $H_{c2}^{\pi}$ determined  in the same way at
more  temperatures  is  displayed  in  Fig.  6. The obtained
temperature dependence reveals  a qualitative agreement with
the   Werthamer,  Helfand   and  Hohenberg   prediction  for
one-band  type-II  superconductor \cite{whh}.

Due  to  the  interband
scattering,  the superconductivity  would be  present in the
$\pi$-band  also above  $H_{c2}^{\pi}$ as  induced from  the
$\sigma$-band until the same $T_c$ and the bulk $H_{c2}$. In
this sense $H_{c2}^{\pi}$ would  only be a certain crossover
field  where  the  small  superconducting  energy  gap  gets
filled. In our data in the  Fig. 5 it is indeed demonstrated
that    the    superconductivity    persists    well   above
$H_{c2}^{\pi}$ and  is still observable at  3-4 T mainly due
to   the   large   energy   gap   $\Delta_{\sigma}$  on  the
$\sigma$-band.

The  effect reveals differently  in
different experiments.  Bouquet et al.  \cite{bouquet2} have
shown in their low  temperature specific heat experiments on
MgB$_2$ single crystals that up  to 0.5 Tesla the electronic
contribution  term  $\gamma$  grows  extremely  rapidly  and
isotropically  with  magnetic  field.  The  effect  has been
attributed  to   the  filling  of  the   small  gap  on  the
$\pi$-band. Then,  at higher fields  anisotropic increase of
$\gamma$ was  observed crossing the  normal state values  at
$H_{c2||c} \approx  $ 3 T and  $H_{c2||ab} \approx 18-22$ T.
By   scanning  tunneling   spectroscopy  Eskildsen   et  al.
\cite{eskildsen}  have  observed  a  very  large vortex core
$\xi^{\pi}  =   50$  nm  for   the  magnetic  field   $H||c$
corresponding  to $H_{c2}^{\pi}$  =  0.13  T while  the real
$H_{c2||c}$ was again about 3 T.

In recent  transport, $ac$-susceptibility and  specific heat
measurements on the MgB$_2$ single crystals \cite{lyard}, we
have shown the temperature  dependence of the upper critical
field for both principal  orientations with $H_{c2||c}$(0 K)
$\simeq $ 3.5 T and $H_{c2||ab}$(0 K) $\simeq 17$ T. In the
point-contact measurements shown in Fig. 5
the superconducting feature is getting completely suppressed
(i.e.  the $\sigma$-band  large gap  is suppressed)  above 5
T indicating  that   this  particular  orientation   of  the
magnetic field with respect to the crystallite under the tip
was in between the $c$-axis and the $ab$-planes of MgB$_2$.

In  Fig. 7  a behavior  of the  Cu-MgB$_2$ junction  with
two superconducting energy gaps as  a function of
the  magnetic  field  is  shown. Data
are presented  for two different contacts  (one data set for
fields below 3 T and the other for higher fields up to 20 T)
with comparable zero-field spectra.  The  weight  parameter
$\alpha $ was equal to 0.68 what is in a perfect agreement
with the  calculations of Brinkman  et al. \cite{brinkman}
for  the important  MgB$_2$  $ab$-plane  current  component.
The small gap is again suppressed
at  small   fields  around  1-2   Tesla  but  the   large  gap
$\Delta_{\sigma}$   persists  close   to  $H_{c2||ab}$.   We
conclude  that  while  the  smaller  gap  is
rapidly closed by  a small magnetic field, the  large gap is
closed at the anisotropic $H_{c2}$.

Similar conclusions  on a rapid filling  of the small energy
gap $\Delta_{\pi}$ in a field of about 1 Tesla and an anisotropic
filling of  $\Delta_{\sigma}$ at much  higher fields near the  real
$H_{c2}$ have  been confirmed by the  later study of Gonnelli
et al. \cite{gonnelli2}.

\section{Electron-phonon interaction}

Ballistic contacts  between normal metals  have been successfully
used  for  a  direct   measurement  of  the  electron-phonon
interaction  \cite{Yanson74,Jansen80}. For  the situation of
a long  mean-free-path  compared  to  the  contact size, the
electrons passing the metallic  contact are accelerated with
a well  defined  excess  energy  $eV$  given  by the applied
voltage  over  the  contact.  After  an inelastic scattering
process (for  instance the spontaneous  emission of phonons)
the injected  electrons can flow  back through the  contact.
These inelastic scattering processes  lead to corrections to
the current  which depend on the  applied voltage because of
the  energy-dependent  scattering  rate  of  the  electrons.
Finally,  the  measured  second  derivative  $d^2V/dI^2$  is
directly  proportional  to  the  electron-phonon interaction
function $\alpha^2F$. The Eliashberg function $\alpha^2F$ is
the convoluted  product of the phonon  density of states $F$ and
the matrix element $\alpha$ for the electron-phonon interaction. This
method has  been extensively employed  for the investigation
of  the   electron-phonon  interaction  in   normal  metals.
However, for superconductors the same  method can be used by
applying a magnetic field to drive the metal into the normal
state.

Point-contact  studies  of  the  non-linear  current-voltage
characteristics  in  the  superconducting  state  have  also
revealed features of the  electron-phonon interaction in the
$d^2V/dI^2$ curves (see for a review \cite{Yanson02}). These
structures are  explained in terms  of either the  inelastic
quasiparticle  scattering  near  the  point  contact  or the
energy-dependence of  the energy gap. The former
mechanism  is similar  to the  point-contact spectroscopy in
the normal  metals: the quasiparticles  generate phonons and
are  back scattered  to the  free quasiparticle  states, now
those  available in  the superconducting  state. The  latter
mechanism   is    of   importance   for    strong   coupling
superconductors.  The  elastic  current  channel  in  the  point
contact comprises
the excess current due to the Andreev reflection and depends
on  the  superconducting  energy  gap,  analogously  to  the
Rowell-McMillan   spectroscopy    of   the   electron-phonon
interaction  in tunneling  experiments. If  the spectrum  is
just  due to  the second  mechanism it  can not  be observed
above the superconducting phase transition.

A few studies have been devoted  on an experimental study of
the  electron-phonon  interaction  in  MgB$_2$  using  point
contacts  \cite{bobrov,D'yachenko,Szabo02,Yanson3}.
The  measurements  of  \cite{D'yachenko}  have  been  done on
Nb-MgB$_2$ (compacted powder)
junctions  at  4.2  K.  The  observed  nonlinearities in the
$d^2V/dI^2(V)$ characteristics have been interpreted
in the frame work of the   Rowell-McMillan
spectroscopy of the electron-phonon interaction in tunneling
experiments accounting for the proximity effect.
The    calculated     Eliashberg    function
reveals  three peaks  located at  37, 62  and 90-92 meV. The
energy position is in agreement with the characteristic
phonon   energies  of MgB$_2$ as   detected   by  neutron  experiments
\cite{Osborn01,Yildrim}.   The  peak intensity at  62 meV does
not show  a prevailing amplitude.
The  calculations  of  Liu  et al.
\cite{liu} predict a strong
anharmonic  electron-phonon  coupling  between  the in-plane
boron mode $E_{2g}$ near  60~~meV and the $p_{x-y}$ orbitals
forming the 2D Fermi sheets, but to resolve this  the
$k$-space  sensitive  experiments  on  single  crystals  are
necessary.

Figure  8 shows   the
$d^2V/dI^2$  spectra of  the point  contact with
significant two gap features.
Now the spectra  are shown in a voltage (energy)  scale up to
120~mV.
They were measured below and  above the
superconducting  transition $T_c(H)$ \cite{Szabo3}.  The  top  panel  shows  $d^2V/dI^2$
curves measured in a fixed magnetic field of 9 Tesla at different
temperatures.  The bottom  panel displays  field dependencies  of
$d^2V/dI^2$ curves at  a fixed temperature of $T$  = 30 K.
The dotted lines in both panels represent
the spectra in the normal state.
At the top curves in the both panels, the
observed  structure   below  10  meV   is  related  to   the
superconducting energy gap.  The
$d^2V/dI^2$  spectra in  the normal  state  show
structures  centered around  38, 62,  80, and  92 mV. Similar
point-contact data  have been reported  by Bobrov et  al. on
$c$-axis oriented MgB$_2$ thin films \cite{bobrov}. These
voltage  positions can  be well  compared with  the observed
peaks  in  the  phonon   density  of  states  obtained  from
inelastic  neutron scattering  experiments (acoustic  phonon
peak  at  36 meV,  optic  bands  at  54,  78,  89  and 97 meV
\cite{Osborn01}).  In  the  point-contact  data the $E_{2g}$
mode is more situated around 60~meV instead of around 54 meV
from  the neutron  data. Both  experimental techniques  show
a broadening of  up to 10  meV in phonon  data which can  be
partly  ascribed  to  the  dispersion  of  the  optic phonon
branches.

The calculations of Liu et al. \cite{liu} beside  a strong
anharmonic  electron-phonon  coupling  between  the
$E_{2g}$ mode and the 2D Fermi sheets, foresee
a hardening (12~\%)  of this mode   upon  entering
into the superconducting state. Choi et al. \cite{choi} have
predicted  the  dominant  peak  in  the  Eliashberg function
$\alpha^2F(\omega)$ at  63 meV for  the case of  harmonic phonons
but at 77 meV for anharmonic phonons.  We      also
do not observe
a particularly strong one peak  in comparison with others in
this energy range in  the point-contact
spectra.
Moreover, there  is no significant change  in position and  intensity of
the  electron-phonon  interaction   structure  upon  passing
through  the superconducting  transition \cite{Szabo02}.  It
means  that observed  structures in  our $d^2V/dI^2$ spectra
are not due to the strong  coupling effect but rather due to
the inelastic scattering.

     Following experimental   studies    of    Yanson    et   al.
\cite{Yanson3}   show   similar energy  correlations   of   their
$d^2V/dI^2(V)$  spectra with  the phonon  density of  states
\cite{renker}.   Moreover,   it   is   shown   that  in  the
point-contact spectra with a small value
of the MgB$_2$ superconducting energy gap ($\Delta_{\pi}$) only a
weak  structure due  to the  electron-phonon interaction  is
observed but  tke structure  grows when  $\Delta$ is increased  to about
3.5 meV presumably  due to the interband scattering.

Raman
studies  have  also  not   observed  any  hardening  of  the
$E_{2g}$   mode   below   the   superconducting   transition
temperature \cite{Martinho01}. We note that the spectroscopic
measurements  of the  electron-phonon interaction  have been
done  on polycrystals  of unknown  contact orientation. Like
for  the  case  of  the  energy-gap  structure,  the contact
orientation    will    also     influence    the    measured
electron-phonon-interaction  spectra   via  the  anisotropy.
Therefore,  single  crystal  studies  are  expected  to give
further   conclusive   information   with   respect  to  the
electron-phonon interaction in MgB$_2$.

\section{Conclusions}

In  conclusion,  we  have  demonstrated  that  the magnesium
diboride is  a two-gap superconductor. The  small gap on the
three-dimensional $\pi$-band is very  weakly coupled  with
$2\Delta_{\pi}/k_BT_c\simeq  $  1.7  and   the  large  gap  is
strongly coupled with
$2\Delta_{\sigma}/k_BT_c \simeq$ 4.  The  small
superconducting   energy  gap   can  be
suppressed by the  magnetic field of about 1-2  Tesla at low
temperatures.  This  field  represents  the $\pi$-band upper
critical  magnetic  field  $H_{c2}^{\pi}$  if  there  was no
interband scattering to the $\sigma$-band. Its temperature
dependence presented for the first time in this paper,
reveals  conventional  behavior  predicted  for  a one-band
type-II superconductor. The
large gap along with the anisotropy of the Fermi surface are
decisive  for  the  real  upper  critical  field of MgB$_2$.
Superconductivity in MgB$_2$ is mainly supplied by electrons
from    the    $\sigma$-band.
Preliminary   point-contact
experiments on the
electron-phonon interaction  reveal the phonon  modes in the
measured   second   derivatives   of   the   current-voltage
characteristics but
do not  show a particular stronger  coupling of the in-plane
$E_{2g}$ boron mode.
Forthcoming  point-contact  experiments  on  single crystals
will  be  very  worthwhile  to  resolve  the  problem  of the
anisotropic  electron-phonon interaction.
\section{Acknowledgments}
This   work  has   been  supported   by  the   Slovak  grant
APVT-51-020102,   the  European Science Foundation ESF -  VORTEX programme  and the
international bilateral programme
of cooperation
between the CNRS and the Slovak Academy of Sciences.

\end{multicols}

\newpage

\widetext
\begin{figure}[tbp] \centerline{ \epsfxsize 6cm \epsffile{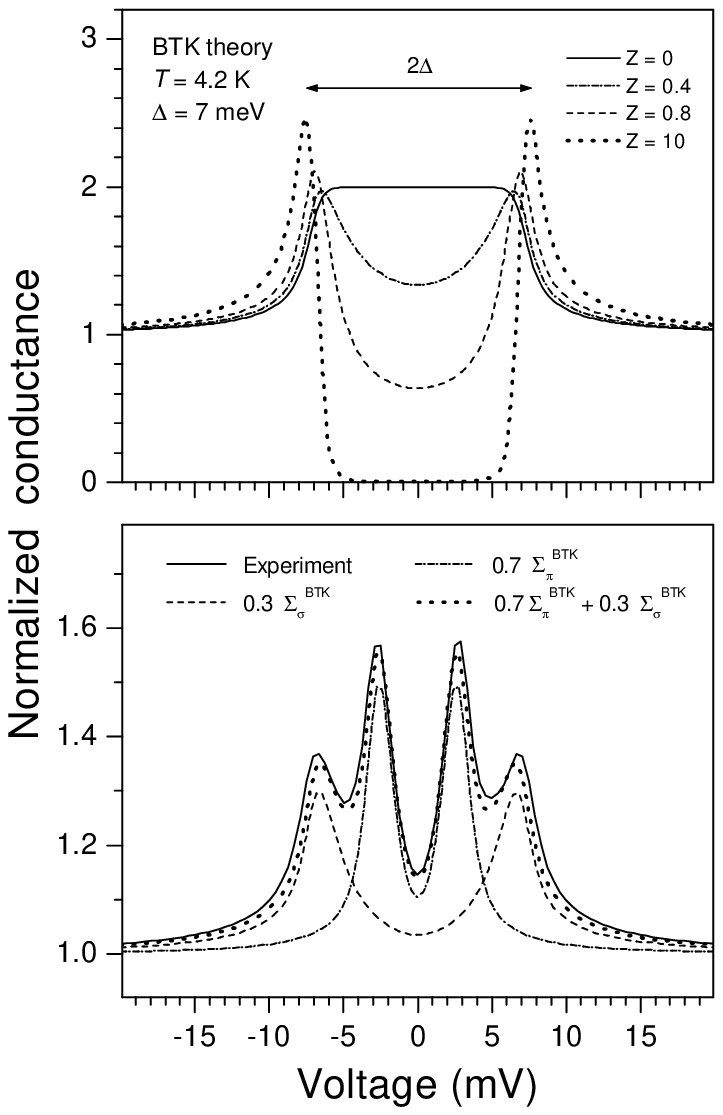}}
\vspace{1cm}
\caption
{a)  Numerical  simulation  of  the  BTK  model  at
different values  of the barrier  strength $Z$, representing
behavior of  the point-contact spectra for  $\Delta =$ 7 meV
between  Giaever  tunneling  ($Z  =$  10)  and  clean Andreev
reflection  ($Z  =  $  0)  at  $T=$  4.2 K [10].
b) The Andreev
reflection spectrum with two
gaps on  the Cu-MgB$ _2$  junction together with  fitting by
two BTK
conductances  (0.3 $\Sigma_{\sigma}$  and 0.7 $\Sigma_{\pi}$
curves are shifted to unity). The final fitting
curve has been generated at
$Z = 0.5$,
$\Delta_{\pi} =  2.8$ meV, $\Delta_{\sigma} =  7 $ meV, $\Gamma
=0 $ with a weight $\alpha = $0.7 at $T = $ 4.2 K [29].}
\end{figure}

\newpage

\begin{figure}[tbp] \centerline{ \epsfxsize 6cm \epsffile{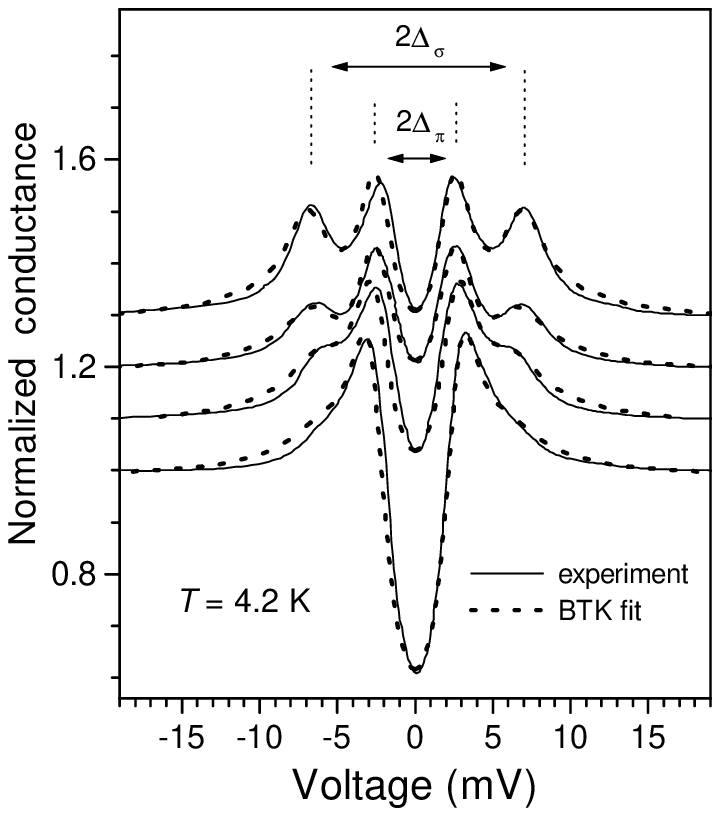}} \vspace{1cm}
\caption{b) Cu - MgB$_2$ point-contact spectra
at $T = 4.2$ K (full lines). The upper curves are vertically
shifted  for  the  clarity.  Dotted  lines  display  fitting
results for  the thermally smeared BTK  model with $\Delta_{\pi}
= 2.8  \pm 0.1$  meV, $\Delta_{\sigma}  =  6.8  \pm 0.3  $ meV  for
different     barrier      transparencies and  weight
factors  [10].}
\end{figure}


\begin{figure} \centerline{ \epsfxsize 6cm \epsffile{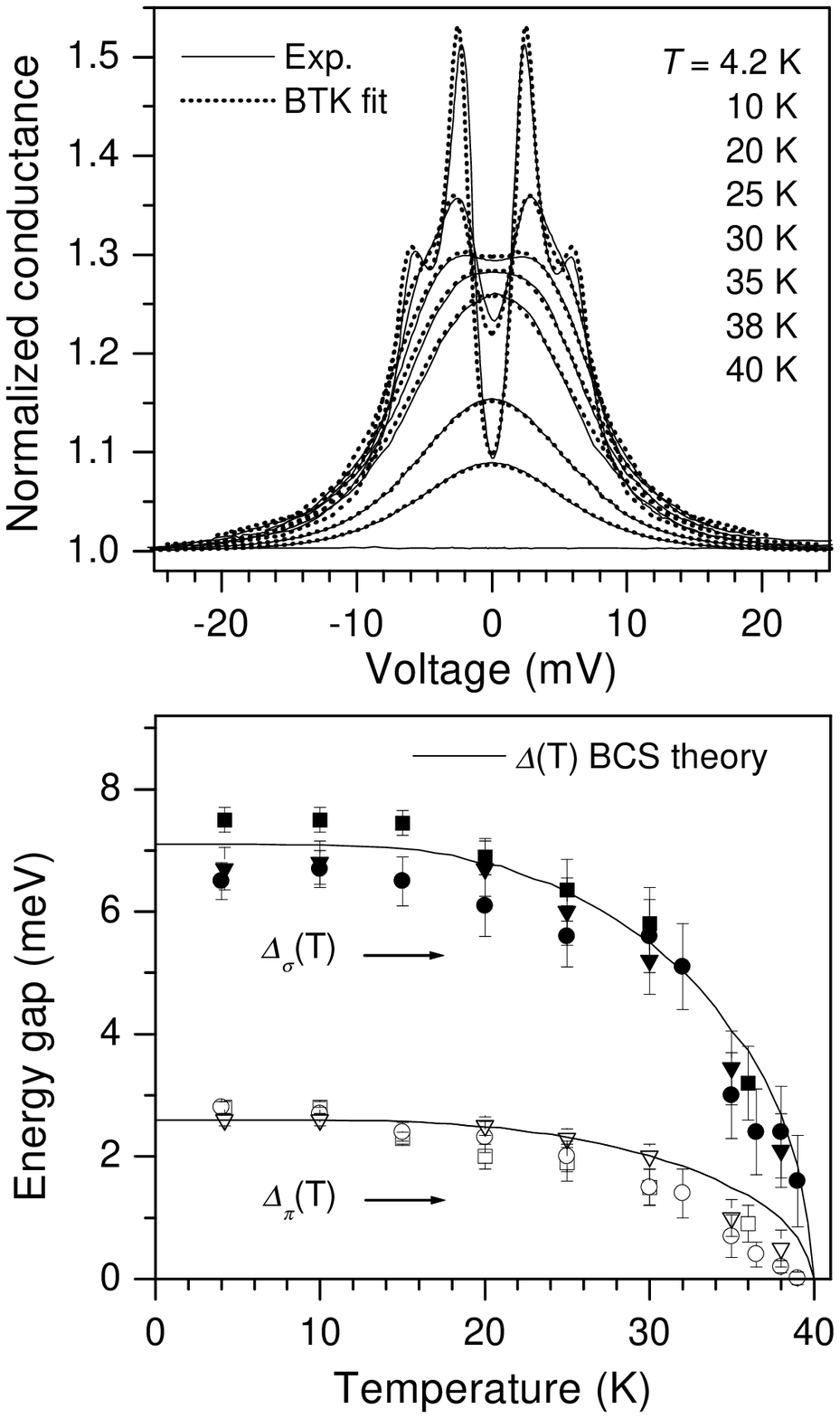}} \vspace{1cm}
\caption{a)  Differential  conductances  of  Cu  -  MgB$_2$
point-contact  measured  (full  lines)  and  fitted  (dotted
lines)  for the  thermally  smeared  BTK model  at indicated
temperatures. The  fitting parameters $\alpha = $ 0.71, $Z =
0.52 \pm 0.02$, $\Gamma = 0.02 $ meV   had the same values at  all temperatures. b)
Temperature dependencies of  both energy gaps ($\Delta_{\pi}(T)$
- solid  symbols, $\Delta_{\sigma}(T)$  - open  symbols) determined
from  the  fitting  on  three  different  point-contacts are
displayed   with  three   corresponding  different  symbols.
$\Delta_{\pi}(T)$ and  $\Delta_{\sigma}(T)$ points determined  from the
same  contact are  plotted with  the same  (open and  solid)
symbols. Full lines represent BCS predictions [10].}
\end{figure}


\begin{figure}       \centerline{       \epsfxsize       8cm
\epsffile{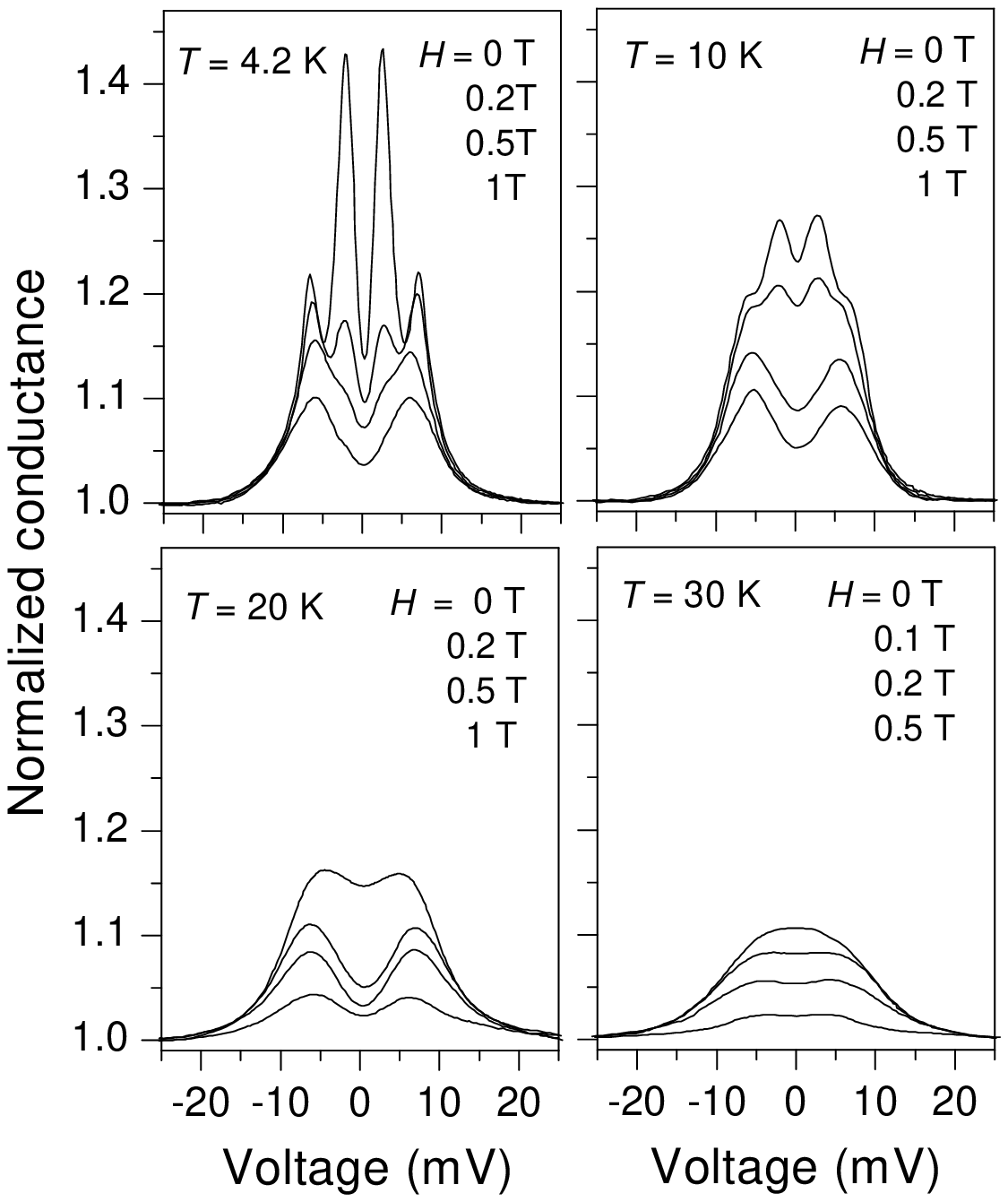}} \vspace{1cm}
\caption{Experimentally  observed influence  of the  applied
magnetic field  on the two-gap structure of  the normalized
point-contact  spectra  at   indicated  temperatures.  These
spectra clearly  reveal that both gaps  exists in zero field
up to $T_c$  as shown by the rapid  suppression of the small
gap  structure ($\Delta_{\pi}  = 2.8$  meV) with  magnetic field
which leads to a broad deep minimum at zero bias  [10].}
\end{figure}

\newpage

\begin{figure}       \centerline{       \epsfxsize      6cm
\epsffile{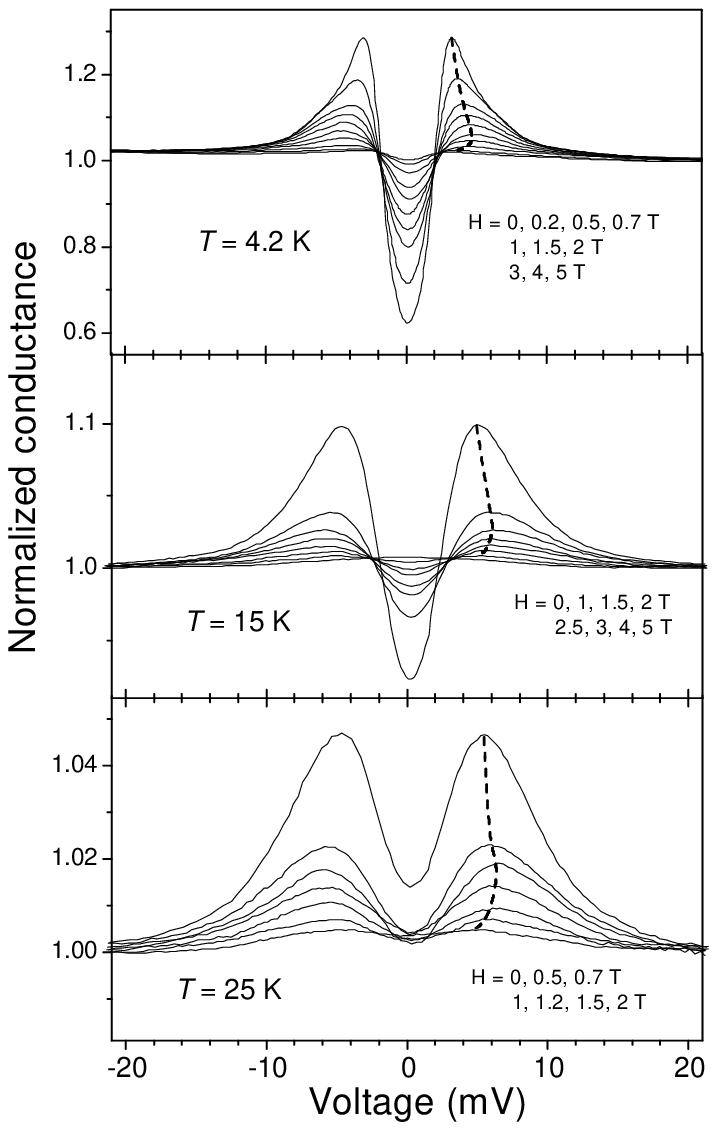}} \vspace{1cm}
\caption{One-gap  spectrum of  the Cu-MgB$_2$  junction with
the  $c$-axis  current injection    at
4.2 K, 15 K and 25 K in different
magnetic fields.
The dashed lines show evolution of the peak position in
magnetic field. The maximal value of this voltage position is a measure of the
critical field for  the suppression of the small  gap.}
\end{figure}

\newpage

\begin{figure} \centerline{ \epsfxsize 6cm \epsffile{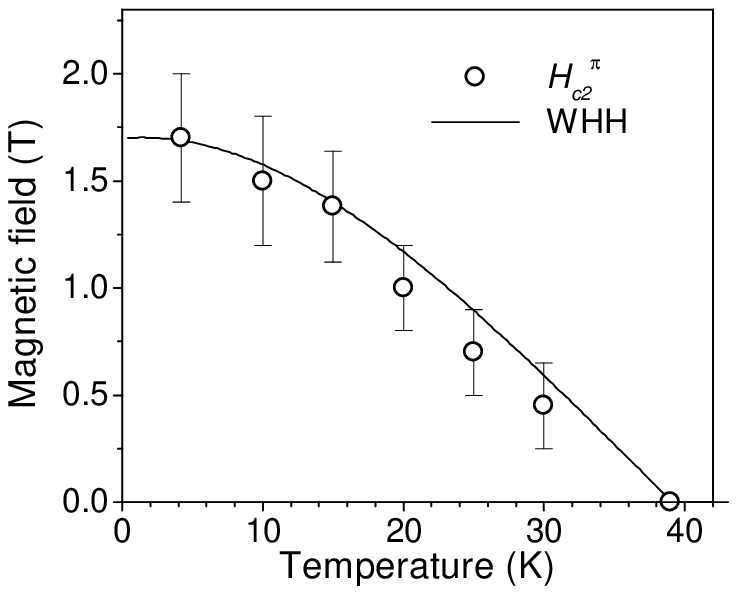}} \vspace{1cm}
\caption{Temperature   dependence   of   the   $\pi$-band  pair-breaking
magnetic field $H_{c2}^{\pi}$}.
\end{figure}

\newpage

\begin{figure}       \centerline{       \epsfxsize      8cm
\epsffile{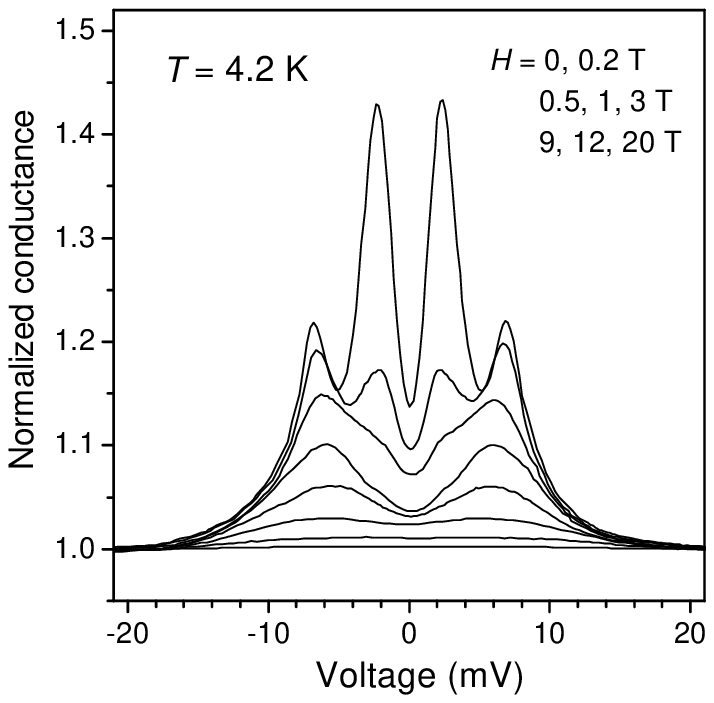}} \vspace{1cm}
\caption{Cu-MgB$_2$ point-contact spectrum  showing two gaps
with an important $ab$-plane current contribution
in  fields  approximately  parallel  to  the  $ab$-planes of
MgB$_2$ up to 20 T at 4.2~K [37].
}
\end{figure}

\newpage

\begin{figure}       \centerline{       \epsfxsize      6cm
\epsffile{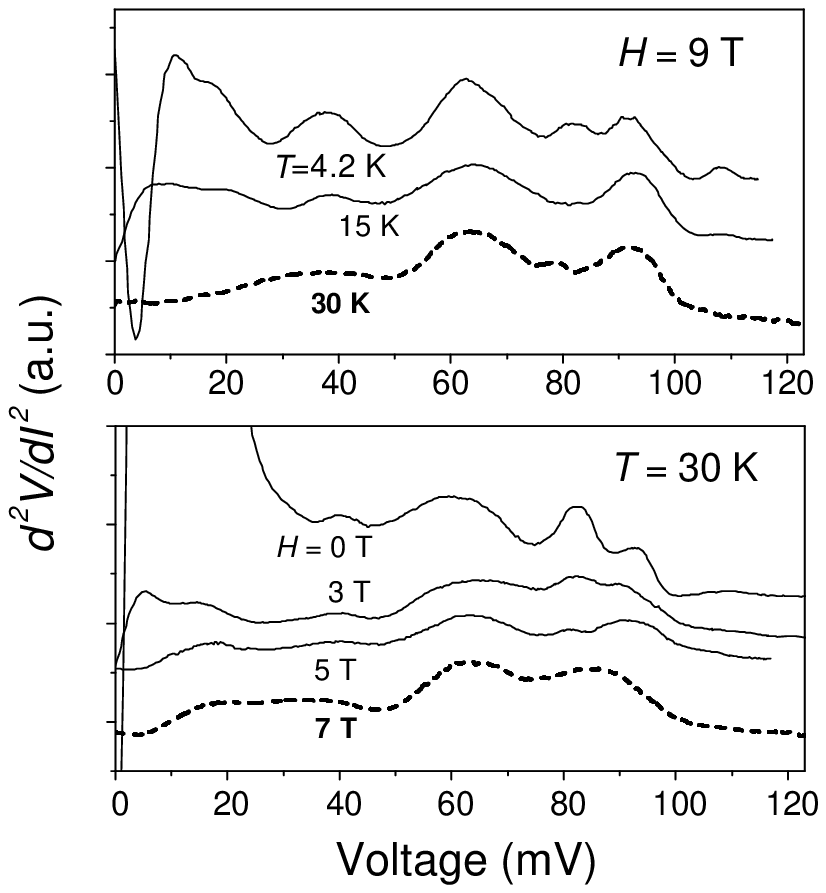}} \vspace{1cm}
\caption{Point-contact   spectra   $d^2V/dI^2(V)$   of   the
Cu-MgB$_2$ junction
measured  at different  temperatures  and magnetic  fields
below  and  above  the  superconducting transition
(dashed lines in the normal state) [16].}
\end{figure}

\begin{references}
\bibitem{nagamatsu} J.  Nagamatsu et al.,  Nature {\bf 410},
63 (2001).
\bibitem{suhl}
H. Suhl, B. T. Matthias, and  L. R. Walker, Phys. Rev. Lett. {\bf 3}, 552 (1959).
\bibitem{binning}
G. Binning, A. Baratoff, H. E.  Hoenig, and J. C. Bednorz, Phys. Rev.
Lett. {\bf 45}, 1352 (1980).
\bibitem{liu}
A. Y. Liu, I. I. Mazin, and J. Kortus, Phys. Rev. Lett. {\bf 87}, 087005 (2001).
\bibitem{bouquet}
F. Bouquet et al., Phys. Rev. Lett. {\bf 87}, 047001 (2001);
Y. Wang et al., Physica C {\bf 355}, 179 (2001).
\bibitem{giubileo}
F.  Giubileo  et  al.,  Phys.  Rev.  Lett.  {\bf 87}, 177008
(2001).
\bibitem{iavarone}
M. Iavarone  et  al., Phys.  Rev.  Lett.  {\bf 89}, 187002
(2002).
\bibitem{suderow}
H. Suderow et al., Physica C {\bf 369}, 106 (2002).
\bibitem{chen}
X. K. Chen et al., Phys.  Rev.  Lett.  {\bf 87}, 157002
(2001).
\bibitem{szabo}
P. Szab\'o et al., Phys. Rev. Lett. {\bf 87}, 137005 (2001).
\bibitem{bouquet2}
F. Bouquet et al.,  cond-mat/0207141.
\bibitem{eskildsen}
M. R. Eskildsen et al., Phys. Rev. Lett. {\bf 89}, 187003.
\bibitem{yelland}
E. A. Yelland et al., Phys. Rev. Lett. {\bf 88}, 217002 (2002).
\bibitem{bobrov}
N.  L. Bobrov  et al.,  in   {\it  New
Trends  in  Superconductivity},  eds.  J  F.  Annett and S.
Kruchinin, Kluwer  Academic Publishers, Dodrecht  2002, NATO
Science Series II: Mathematics,  Physics and Chemistry, Vol.
67, p. 225.
\bibitem{D'yachenko}    A.    I.     D'yachenko    et    al.,
cond-mat/0201200.
\bibitem{Szabo02}
P.  Szab\'o, P.  Samuely, J.  Ka\v cmar\v  cik, T. Klein, J.
Marcus, and  A. G. M. Jansen,  to appear in  Physica C.
\bibitem{Yanson3}
I. K. Yanson et al., cond-mat/0206170.
\bibitem{mullen}
K. Mullen, E.  Ben-Jacob and S. Ruggiero, Phys.  Rev. B {\bf
38}, 5150 (1988).
\bibitem{Blonder82}
G. E. Blonder, M. Tinkham, and T. M. Klapwijk, Phys. Rev. B {\bf 25},
4515 (1982).
\bibitem{plecenik}
A.Plecenik et al.,  Phys.  Rev. B {\bf 49}, 10016 (1996).
\bibitem{schmidt}
H. Schmidt, J. F. Zasadzinski, K. E.  Gray, and D. G. Hinks,
Phys. Rev. B {\bf 63}, 220504 (2001).
\bibitem{kohen}
A.  Kohen and  G.  Deutscher, Phys.  Rev. B  {\bf 64},  060506(R)
(2001).
\bibitem{kortus}
J. Kortus et al., Phys. Rev. Lett. {\bf 86}, 4656 (2001).
\bibitem{plecenik1}
A. Plecenik, \v  S. Be\v na\v cka, P.  K\'u\v s, and M. Grajcar,
Physica C {\bf 386}, 251 (2002).
\bibitem{laube}
F. Laube et al., Europhys. Lett. {\bf 56}, 296 (2001).
\bibitem{gonnelli}
R. S.  Gonnelli et al.,  Int. J. Mod.  Phys. {\bf 16},  1553
(2002).
\bibitem{lee}
S. Lee et al., Physica C {\bf 377}, 202 (2002).
\bibitem{li}
Z.-Z. Li et al., Phys. Rev. B {\bf 66}, 064513  (2002).
\bibitem{ttpam}
P.  Szab\'o, P.  Samuely, J.  Ka\v cmar\v  cik, T. Klein, J.
Marcus,  A. G. M. Jansen,  to appear in Physica B (2003).
\bibitem{brinkman}
A. Brinkman et al., Phys. Rev. B {\bf 65}, 180517(R) (2001).
\bibitem{bugoslav}
Y. Bugoslavsky  et al., Supercond.  Sci. Technol. {\bf  15},
526 (2002).
\bibitem{naidyuk}
Yu. G. Naidyuk et al., JETP Letter {\bf 75}, 238  (2002).
\bibitem{gonnelli2}
R. S. Gonnelli et al., cond-mat/0208060.
\bibitem{BPBO}
P. Szab\'o et al., J. Low Temp. Phys. {\bf 106}, 291 (1997);
P.  Szab\'o, P.  Samuely, A.   G. M.  Jansen, J.  Marcus, P.
Wyder, Phys. Rev. B {\bf 62}, 3502 (2000).
\bibitem{whh}
N.R.Werthamer, E.Helfand, and P.Hohenberg, Phys. Rev. {\bf 147}, 295 (1966).
\bibitem{lyard}
L. Lyard et al., to appear in Phys. Rev. B (2002); cond-mat/0206231.
\bibitem{boromag}
P.  Szab\'o  et  al.,  to appear in  Supercond. Sci.
Technol.
\bibitem{Yanson74}
I.K. Yanson, Zh. Eksp. Teor. Fiz. {\bf 66}, 1035 (1974)
[Sov. Phys. - JETP {\bf 39}, 506 (1974)].
\bibitem{Jansen80}
A.G.M. Jansen, A.P. van Gelder, and P. Wyder,
J. Phys. C: Solid State Phys. {\bf 13}, 6073 (1980).
\bibitem{Yanson02}
I.K. Yanson, in
{\it Quantum Mesoscopic Phenonema  and Mesoscopic Devices in
Microelectronics},  eds. I.  O. Kulik  and R.  Ellialtioglu,
 Kluwer Academic Publishers, 2000, p.61.
\bibitem{Osborn01}
R. Osborn et al., Phys. Rev. Lett. {\bf 87}, 017005 (2001).
\bibitem{Yildrim}
T. Yildrim et al., Phys. Rev. Lett. {\bf 87}, 037001 (2001).
\bibitem{Szabo3}
P. Szab\'o et al., Physica C {\bf 369}, 250 (2002).
\bibitem{choi}
H. J. Choi, D. Roundy, H. Sun, M. L. Cohen, and S. G. Louie,
Phys. Rev. B {\bf 66}, 020513 (R) (2002).
\bibitem{renker}
B. Renker et al., Phys. Rev. Lett. {\bf 88}, 067001 (2002).
\bibitem{Martinho01}
H. Martinho et al., cond-mat/0105204.

\end{references}
\end{document}